\documentclass{mem}
\usepackage{natbib}\usepackage{txfonts}\usepackage{balance}
\usepackage{graphicx}
\usepackage[a4paper,breaklinks,dvipdfm]{hyperref}
\idline{0}{0}
\begin{document}
\def\teff{$T\rm_{eff }$}
\def\kms{$\mathrm {km s}^{-1}$}
\def\Swift{{\it Swift}}
\def\Mo{M_\odot}

\title{
Short gamma-ray bursts: evidence for an origin in globular clusters?
}

   \subtitle{}

\author{
R.~P.~Church\inst{1},
A.~J.~Levan\inst{1,2}
M.~B.~Davies\inst{1}
\and N.~Tanvir\inst{3}
          }

  \offprints{R.~P.~Church}

\institute{
Department of Astronomy and Theoretical Physics,
Lund Observatory,
Box 43,
SE 221 00 Lund,
Sweden
\and
Department of Physics, University of Warwick, Coventry. CV4 7AL. UK
\and
Department of Physics and Astronomy, University of Leicester, Leicester.  KE1
7RH.  UK\\
\email{ross@astro.lu.se}
}

\authorrunning{Church et al.}

\titlerunning{SGRBs from globular clusters}

\abstract{
We compare the observed spatial offsets of short gamma-ray bursts from their host
galaxies with their predicted distributions, assuming that they originate in
double neutron star binaries that form from field stars.  We find that, for the
majority of bursts, this model is sufficient to explain the observed offsets,
although there is a trend towards larger offsets
than predicted.  One burst, GRB~060502B, has an offset that is clearly
anomalous.  We discuss possible reasons for the large offsets, including host
galaxy misidentification, and suggest that some of the largest-offset bursts may
originate in the merger of double neutron star binaries that form dynamically in
the cores of globular clusters.
}
\maketitle{}

\section{Introduction}

The advent of the \Swift{} satellite has, for the first time, allowed
us to locate the positions on the sky of short-duration gamma-ray bursts (SGRBs)
and hence determine their host galaxies.  The sample of located short bursts
that has built up over the last six years of \Swift{} operations covers a 
range of galaxy types, including elliptical galaxies in which no stars are
currently being formed.  The bursts are observed to occur with a wide range of
spatial offsets from their inferred host galaxies, in some cases occurring well
outside the host.

These observations broadly support a picture in which the bursts are powered by
the formation of a stellar-mass black hole during the merger of a binary
containing either two neutron stars (NS--NS) or a black hole and a neutron star
(BH--NS).  These mergers are driven by inspiral caused by the emission of
gravitational radiation; this leads naturally to a very wide range of merger
timescales.  Hence SGRBs can occur in non-star-forming hosts.  Meanwhile, the
kicks expected to be present at the formation of the neutron stars will impart a
natal velocity to the binaries, which offers a natural qualitative explanation
of the offsets \citep[e.g.][]{bloom99,fryer99}.

Utilising the hosts identified by \Swift{}, we build on the work of previous
authors by considering the bursts on a
host-by-host basis.  We construct offset distributions based on the properties of
the observed hosts.  We use a sample of 16 bursts, all of which have identified
hosts with measured properties including magnitude and redshift.  
Our burst sample is presented in Table~\ref{tab}.

\begin{table}
\begin{tabular}{llllll}
\hline
GRB     &$R_{\rm off}$ &$v_{\rm h}$& $r_{\rm h}$ &  $R_{\rm e}$   \\
        & kpc          &km/s       & ${\rm kpc}$ &  ${\rm kpc}$   \\
\hline
050509B & 63.7 $\pm$  12.1                & 664 &   46.3 &   21.0 \\
050709  & 3.55 $\pm$  0.27                & 110 &    7.9 &   1.8  \\
050724  & 2.54 $\pm$  0.08                & 532 &   23.9 &   4.0   \\
051221A & 1.53 $\pm$  0.31                & 157 &   15.7 &   2.2  \\
060502B & 73   $\pm$  13                  & 505 &   20.5 &   10.5  \\
060801  & 19.7 $\pm$  14.0                & 170 &   18.2 &   3.0  \\
061006  & 1.44 $\pm$  0.29                & 124 &    9.9 &   3.7  \\
061201  & 33.9 $\pm$  0.4                 & 121 &    9.6 &   1.8  \\
061210  & 10.7 $\pm$  6.9                 & 162 &   16.5 &   2.6  \\
061217  & 55   $\pm$  20                  & 141 &   12.8 &   1.8  \\
070429B & 4.7  $\pm$  4.7                 & 149 &   14.2 &   2.1  \\
070714B & 3.08 $\pm$  0.47                & 111 &    8.9 &   0.94  \\
070724A & 4.76 $\pm$  0.06                & 435 &   13.1 &   3.2  \\
070809  & 19.61$\pm$  1.9                 & 110 &    8.0 &   0.92  \\
071227  & 16.1 $\pm$  0.2                 & 173 &   18.8 &   3.1  \\
080905A & 18.11$\pm$  0.42                & 170 &   18.1 &   3.0  \\
\hline
\end{tabular}
\caption{Properties of SGRBs in our sample and their host Galaxies.  For more
details of the population and a complete reference list see \citep{church11}.
}
\label{tab}
\end{table}

\section{Data and calculations}

\begin{figure}
\includegraphics[width=\columnwidth]{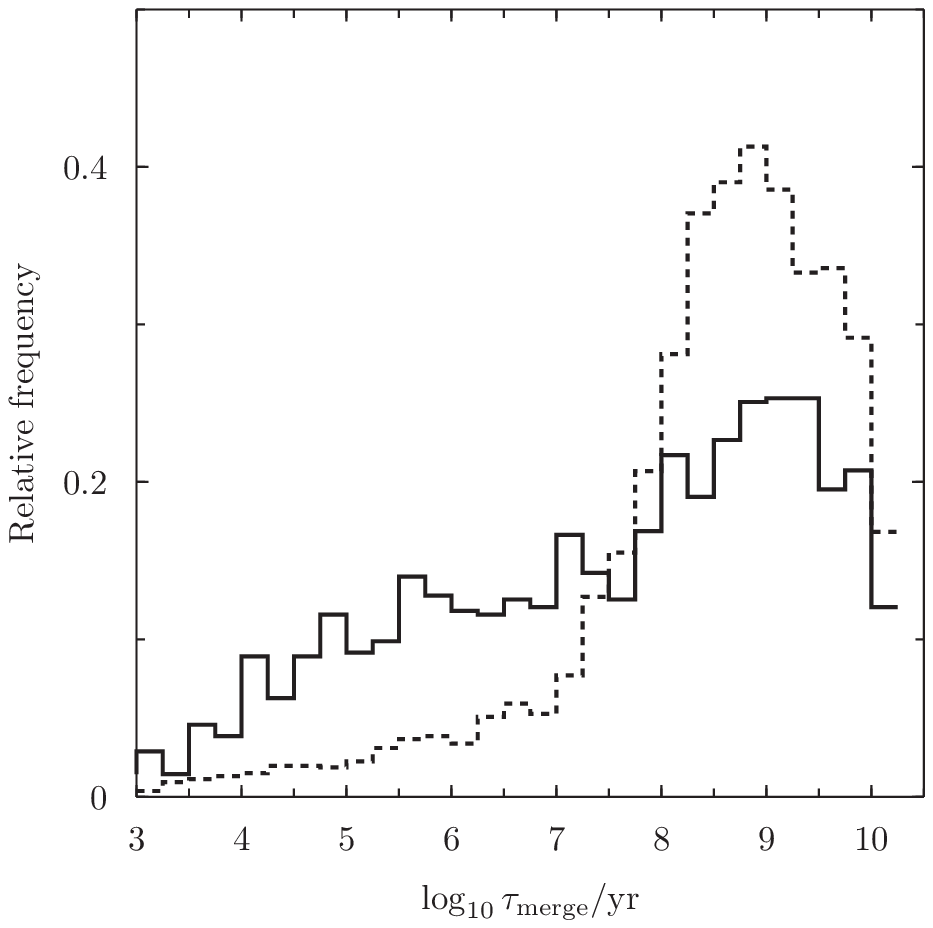}
\caption{\footnotesize
The merger time distribution of our synthesised binary population.
NS--NS binaries are shown in as solid lines; BH--NS binaries as dashed lines.}
\label{fig:mergdist}
\end{figure}

\begin{figure}
\includegraphics[width=\columnwidth]{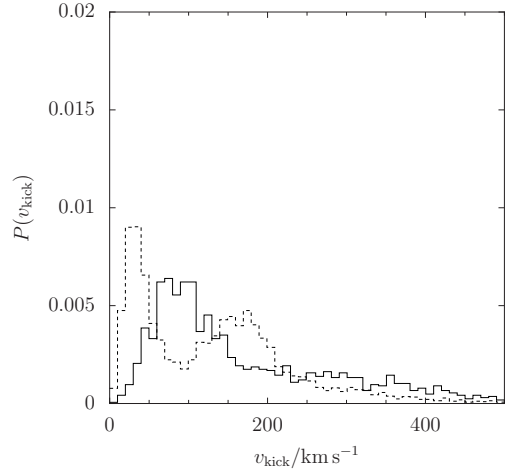}
\caption{\footnotesize
The velocity distribution of our synthesised binary population after the
formation of a double compact binary.  NS--NS binaries are shown in as solid
lines; BH--NS binaries as dashed lines.  Velocities are measured with respect to
the initial rest frames of the binaries.}
\label{fig:veldist}
\end{figure}

In order to predict the offsets of short gamma-ray bursts we synthesised a
large population of compact binaries using the rapid binary evolution code
BSE~\citep{hurley2002}.  We retained those binaries that evolved into NS--NS or
BH--NS binaries, and computed their final 3D velocities, taking into account the
effects of supernova mass loss and natal kicks.  We found that, in order to
satisfactorily reproduce the orbital properties of Galactic NS--NS binaries, we
required strong natal kicks, of the order of $100\,{\rm km\,s^{-1}}$.  Hence we
adopt the \citet{acc02} kick distribution.  We present distributions of merger
times and rest-frame velocities for our final sample of binaries in
Figures~\ref{fig:mergdist}~and~\ref{fig:veldist}.

We model the observed SGRB sample on a burst-by-burst basis.  For each host
galaxy we produce a separate potential model, utilising the logarithmic
potential of \citet{thomas09}.
The core radius $r_{\rm h}$ and halo circular velocity $v_{\rm h}$ for each
burst are obtained from the the fits of \citet{Kormendy04} to SDSS data.  Fits
from the same source were also used for half-light radii where measured values
were not available in the literature.  The properties of our haloes are given in
Table~\ref{tab}.  We form stars in an exponential disc with radial scale length
equal to the half-light radius of the host.  We place our compact binaries in
these discs, endowing them with the calculated velocities in isotropically
distributed directions.  We integrate the motion of each binary's centre of mass
in the potential field of the host for the merger time of the binary, recording
the position in the host galaxy where the merger occurs.  Finally we project the
location onto the host using a random viewing angle. This process is repeated
over a large number of realisations to build up an offset distribution for each
host.  

\begin{figure}
\includegraphics[width=\columnwidth]{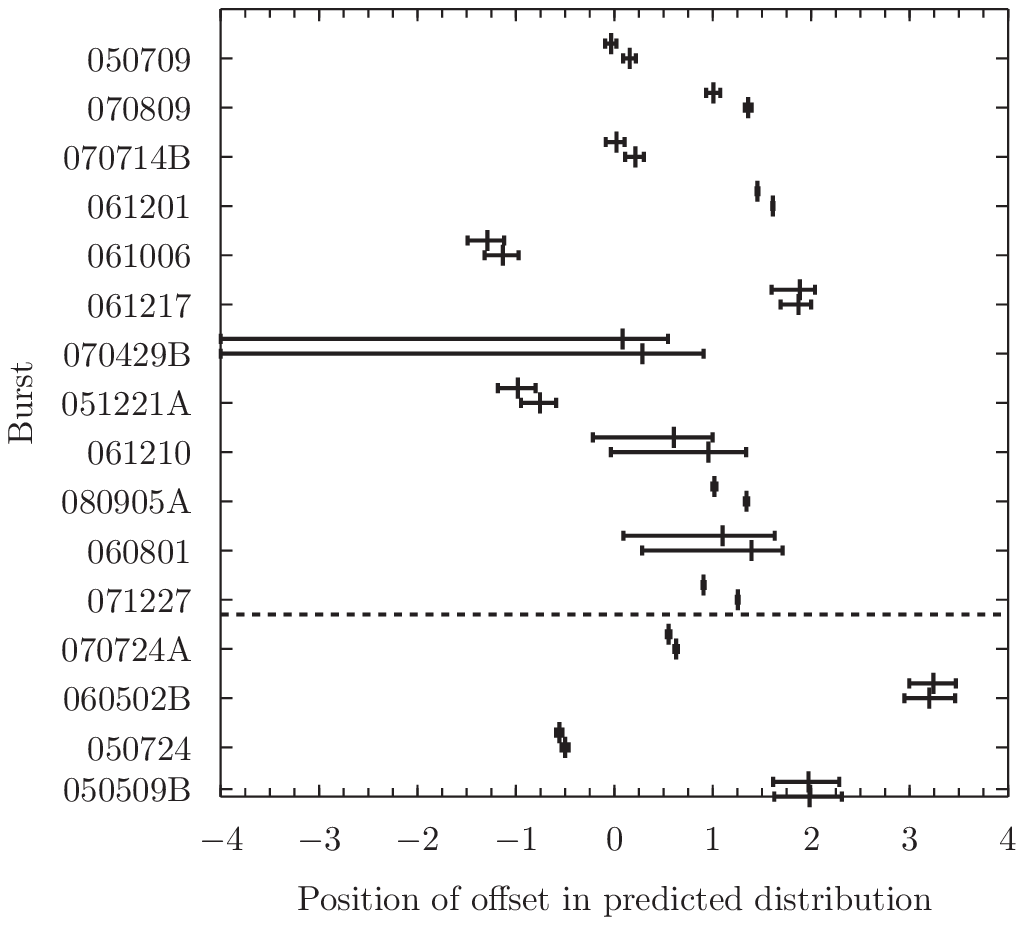}
\caption{\footnotesize
The observed offset from the host centre for each of the bursts in our sample.
To allow the consistent comparison of different hosts we have plotted the
difference between the observed offset and the median predicted offset, in units
of the standard deviation of its host's predicted offset distribution.
For example, a burst which lay at the 98th percentile in the cumulative offset
distribution of its host would be plotted at $x=2$.  If the predicted
distributions match the observed values then the plot should be symmetric around
$x=0$.  For each burst the range plotted is one standard deviation in observed
offset.  The lower bar in each case is for NS--NS progenitors, the upper for
NS--BH progenitors.
}
\label{fig:offsetGaussianPlot}
\end{figure}

\begin{figure}
\includegraphics[width=\columnwidth]{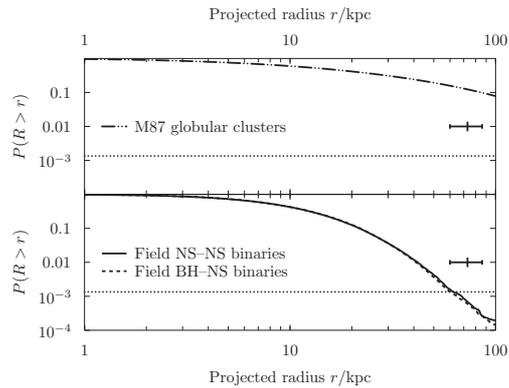}
\caption{\footnotesize
Lower panel: the cumulative offset probability for bursts around the host of
GRB~060502B assuming that the burst comes from a merging field binary.
Solid lines show the distributions for bursts
from NS--NS binaries; dashed lines show bursts from BH--NS binaries.
The error bar, placed at arbitrary height, shows the
1-$\sigma$ error on the measured offset.  The dotted line shows the 3-$\sigma$
exclusion level.
Upper panel: ditto, except for NS--NS binaries that form dynamically in a
globular cluster system similar to that around M87.
}
\label{figCumProb}
\end{figure}

The position of each observed burst within the offset distribution predicted for
its host is plotted in Figure~\ref{fig:offsetGaussianPlot}.  The majority of
the bursts are reproduced relatively well by our treatment, although
it is evident that the synthesised distributions are systematically
under-predicting the host offsets; i.e. the distribution is not centred around
$x=0$.  In particular one burst, GRB~060502B, is at an anomalously large offset.
The cumulative distribution of predicted offsets for this burst is shown in
Figure~\ref{figCumProb}.  The possibility of a burst occurring at such a large
offset around this host galaxy is excluded at the three-sigma level, although
the errors on the burst position are rather large as no optical afterglow was
detected.

\section{Discussion}

There are several possible explanations for the failure of the field binary
scenario to reproduce the large offset of GRB~060502B.  Firstly, it is possible
that the host galaxy has been mis-identified.  We follow \citet{bloom2007} in
selecting the giant elliptical considered here as the host with a large offset.
The probability of a chance co-incidence with this galaxy is only a few per
cent; on the other hand there are smaller, fainter galaxies within the XRT error
circle \citep{bergerhiz}.  This is an innate problem in measuring offsets with
hosts selected through spatial co-incidence on the sky.

A second possibility is that, for bursts around massive elliptical galaxies, the
evolution of the hosts has driven the binaries out to larger offsets.
\citet{zemp09} compute the distribution of coalescing compact binaries in
cosmologically evolving dark matter halos and show that the result is to
increase the observed offsets, with the inferred host galaxy not always being
the galaxy in which the binary formed.  This is a entirely possible origin for
the progenitor of GRB~060502B.

The third possibility we consider is that some fraction of bursts
originate in compact binaries that have formed dynamically within the globular
cluster systems of their host galaxies.  Within the cores of globular clusters
the number densities are high enough that stars undergo dynamical
encounters with one another.  Neutron stars, which segregate to the centre of
the clusters, can exchange into pre-existing binaries to form NS--NS binaries
\citep{davies1995}.  Those binaries then could yield SGRBs.
\citet{grindlay2005} extrapolate from M15-C, the single NS--NS binary that is
unambiguously dynamically formed, to derive a merger rate of $R_{\rm
gc}=0.31T(10^9\,\Mo)^{-1}t_{\rm Hubble}^{-1}$, where $T$ is the number of
globular clusters per $10^9\,\Mo$ of galactic stellar mass.  They caution,
however, that simple scattering calculations suggest that this rate is an
underestimate by a factor of at least ten.  Our population synthesis implies a 
field NS--NS merger rate of $R_{\rm field}=15f_{\rm b}(10^9\,\Mo)^{-1}t_{\rm
Hubble}^{-1}$, where $f_{\rm b}$ is the field binary fraction.  This suggests
that the ratio of rates of field and globular cluster mergers is
$\frac{R_{\rm gc}}{R_{\rm field}} \simeq 0.02T/f_{\rm b}$.
The nearest elliptical galaxy of similar size to the putative host of
GRB~060502B is M87, which has roughly 14\,000 globular clusters~\citep{harris09}
in a very extended distribution.  Its specific density of globular clusters is
$T\simeq 8$ \citep{Brodie06}.  Taking into account the underestimate mentioned
above and the uncertainty in binary population synthesis
calculations this is consistent with a similar rate of burst production in the
field and in globular clusters for massive elliptical galaxies.  In the upper
panel of Figure~\ref{figCumProb} we plot the cumulative offset distribution
given by assuming that GRB~060502B occurred in a globular cluster in a system
identical to that of M87.  This gives a much better fit to the offset than a
field population origin.

\begin{acknowledgements}
RPC is funded by a Marie-Curie Intra-European Fellowship, Grant No.~252431 under
the European Comission's FP7 framework.  This work was supported by the Swedish
Research Council (grant 2008--4089).  This work utilises computers purchased
with the support of the Royal Physiographic Society of Lund.
\end{acknowledgements}

\bibliographystyle{aa}
\bibliography{new.bib}

\end{document}